\documentclass[prd,preprint,superscriptaddress,amsmath,amssymb,nofootinbib]{revtex4}
\usepackage{graphicx}
\usepackage{dcolumn}
\usepackage{bm}
\usepackage{amssymb}
\usepackage{amsmath}
\usepackage{epsfig}    
\usepackage{color}
\usepackage{slashed}
\usepackage{hhline}

\def\be{\begin{equation}}
\def\ee{\end{equation}}
\newcommand{\bea}{\begin{eqnarray}}
\newcommand{\eea}{\end{eqnarray}}

\begin{document}

\title{Radiative inverse seesaw model with hidden $U(1)$ gauge symmetry enhancing lepton $g-2$}

\author{Takaaki Nomura}
\email{nomura@scu.edu.cn}
\affiliation{College of Physics, Sichuan University, Chengdu 610065, China}

\author{Hiroshi Okada}
\email{hiroshi3okada@htu.edu.cn}
\affiliation{Department of Physics, Kyushu University, 744 Motooka, Nishi-ku, Fukuoka, 819-0395, Japan}
\affiliation{Department of Physics, Henan Normal University, Xinxiang 453007, China}

\date{\today}

\begin{abstract}
We propose a new inverse seesaw model based on hidden local $U(1)$ symmetry framework where 
inverse seesaw mechanism is induced at one loop level.
A Majorana mass term of singlet fermion is forbidden by the $U(1)$ symmetry and it is generated at one-loop level by introducing relevant particle contents to get loop diagram, 
inducing inverse seesaw mechanism.
The same particle contents also contribute to lepton magnetic(electric) dipole moment and lepton flavor violating decays without chiral suppression.
We can then obtain sizable muon anomalous magnetic dipole moment that accommodate with deviation from the standard model prediction.
The constraints from lepton flavor violating decays and electron magnetic(electric) dipole moment are also discussed to explore testability of the model. 

 \end{abstract}
\maketitle

\section{Introduction}

A mechanism to generate tiny neutrino masses is one of the important issue that definitely requires physics beyond the standard model (SM).
A well known scenario is the type-I seesaw mechanism~\cite{Yanagida:1979gs, Minkowski:1977sc, Mohapatra:1979ia} that introduces heavy right-handed neutrinos $(N_R)$ in realizing smallness of active neutrino masses.
Typically  required new physics scale is very high such as around Grand Unified Theory (GUT) one ($\mathcal{O}(10^{16})$ GeV) which is also motivated to work the leptogensis mechanism well~\cite{Fukugita:1986hr}.
In this framework it is thus difficult to see effects of new physics at low energy scale through, for example, lepton flavor violations (LFVs) and electric(magnetic) dipole moments of charged leptons.

An inverse seesaw (IS)~\cite{Mohapatra:1986bd, Wyler:1982dd} mechanism is an alternative scenario to generate the neutrino mass that tends to require lower mass scale compared to type-I seesaw enabling us to obtain higher verifiability. In this case we introduce SM singlet left-handed chiral fermions $(N_L)$ in addition to right-handed neutrinos and active neutrino mass is given by $m_\nu \sim \mu\left(\frac{M_D}{M_N}\right)^2$ where $M_D$ is Dirac mass between active and right-handed neutrinos, $M_N$ is Dirac mass between $N_R$ and $N_L$, and $\mu$ is Majorana mass parameter for $\bar{N}^c_L N_L$ term
violating lepton number. Then neutrino mass can be tiny without much high energy scale if $\mu$ is very suppressed.
Basically one can assume $\mu$ is small as it is related to lepton number violation, 
but we can also consider a mechanism to get its smallness where a Majorana term with $\mu$ is generated at loop level. 
In fact we can realize loop induced inverse seesaw introducing hidden $U(1)$ gauge symmetry where SM particles are not charged under it~\cite{Nomura:2021adf}.
Two singlets $N_R$ and $N_L$ are originated from one vector-like fermion with hidden $U(1)$ charge where it should be vector-like to cancel gauge anomaly.
Then this symmetry plays a role of forbidding Majorana mass term at tree level~\footnote{ Although we do not discuss in this paper, a hidden $U(1)$ gauge symmetry is also motivated for dark matter (DM) physics providing its stability and mediator of interactions with SM particles.~\cite{Zhang:2009dd, Chiang:2013kqa, Chen:2015nea, Chen:2015dea, Gross:2015cwa, Hambye:2008bq, Boehm:2014bia, Baek:2013dwa, Khoze:2014woa, Daido:2019tbm, Karam:2015jta, Davoudiasl:2013jma, Ko:2020qlt, Nomura:2020zlm, Cai:2018upp, Nomura:2017wxf,Matsui:2023bwa, Li:2023fzv, Nagao:2022osm} }. 
After spontaneous symmetry breaking, a Majorana mass term of $N_L$ can be induced at loop level by introducing relevant particle contents to make a loop diagram.
Interestingly these new particle contents also induce LFV and contribute to electric(magnetic) dipole moments.

A test of new physics by lepton flavor physics is achieving high precision and we expect further improvements in near future.
{ In particular muon anomalous magnetic dipole moment (muon $g-2$) would be good test of beyond the SM
where the discrepancy with the SM prediction is 5.1$\sigma$ level estimated by data-driven dispersion relation method~\cite{Muong-2:2021ojo, Muong-2:2023cdq, Muong-2:2006rrc, Aoyama:2012wk,Aoyama:2019ryr,Czarnecki:2002nt,Gnendiger:2013pva,Davier:2017zfy,Keshavarzi:2018mgv,Colangelo:2018mtw,Hoferichter:2019mqg,Davier:2019can,Keshavarzi:2019abf,Kurz:2014wya,Melnikov:2003xd,Masjuan:2017tvw,Colangelo:2017fiz,Hoferichter:2018kwz,Gerardin:2019vio,Bijnens:2019ghy,Colangelo:2019uex,Blum:2019ugy,Colangelo:2014qya,Hagiwara:2011af} 
combining new results from the E989 collaboration at Fermilab~\cite{Muong-2:2021ojo, Muong-2:2023cdq} and the previous result from BNL experiment~\cite{Muong-2:2006rrc} as 
\begin{align}
\Delta a^{\rm data-driven}_\mu = (24.9 \pm 4.9)\times 10^{-10}.
\label{exp_dmu}
\end{align}
In fact we have alternative hadron vacuum polarization (HVP) values, estimated by lattice calculations~\cite{Borsanyi:2020mff,ExtendedTwistedMass:2022jpw,Ce:2022kxy},  shown in refs.~\cite{Crivellin:2020zul,deRafael:2020uif,Keshavarzi:2020bfy}~\footnote{
The lattice results would imply new tensions with the HVP extracted from $e^+ e^-$ data and the global fits to the electroweak precision observables.
The effect in modifying HVP for muon $g-2$ and electroweak precision test is also discussed previously in ref.~\cite{Passera:2008jk}.} that give us muon $g-2$ consistent with the SM. 
The recent White Paper 2025 (WP25) summarizes the muon $g-2$ value as~\cite{Aliberti:2025beg}
\begin{equation}
\Delta a_\mu^{\rm WP25} = (39 \pm 64) \times 10^{-11}.
\end{equation}
 }
We expect more precise test of muon $g-2$ can be a key to understand new physics.
It is thus interesting to consider new physics that can have sizable muon $g-2$. 
To obtain the sizable muon $g-2$ with natural manner by Yukawa couplings we would need one-loop contributions with chiral flip by heavy fermion mass inside a loop diagram~\cite{Lindner:2016bgg,Chen:2016dip, Athron:2021iuf,Guedes:2022cfy,Crivellin:2021rbq} or the Yukawa couplings would exceed perturbation limit.
In this work we explore the possibility to get such a mechanism to enhance muon $g-2$ in a neutrino mass model, especially inverse seesaw mechanism.
In addition to muon $g-2$, this enhancement affects electron $g-2$, LFVs, electric dipole moment(EDM) of electron/muon that are also tested with high precision~\cite{MEGII:2023ltw, MEG:2016leq, BaBar:2009hkt,Renga:2018fpd, Parker:2018vye, Morel:2020dww, Roussy:2022cmp, Muong-2:2008ebm} and 
we check these constraints.

In this paper, we propose a new model with hidden $U(1)$ gauge symmetry where inverse seesaw mechanism is induced at one-loop level by 
introducing some new field contents. 
It is found that our field contents can also enhance lepton ($g-2$) without chiral suppression compared to the previous model~\cite{Nomura:2021adf} and we can explain muon $g-2$ anomaly.
The same enhancements appear in LFV decay amplitude and electron/muon EDMs and the model would be testable in future experiments.
We formulate active neutrino mass, LFV and electric(magnetic) dipole moments in the model, and perform numerical analysis searching for parameter region 
that is consistent with current experimental data of neutrino masses/mixings and lepton flavor constraints.
Then we investigate flavor observables such as LFV decay ratios under the condition where we can obtain sizable muon $g-2$ to show testability of the model.

This paper is organized as follows. In Section~II, we present the model with mass spectrum, relevant interactions and formulas for phenomenology of our interest.
In Section III, we perform numerical analysis to search for allowed parameter region and predicted flavor observables. 
Finally we devote Section~IV to the summary and conclusion.

\section{A model}

\begin{table}[t]
  \begin{center}
    \begin{tabular}{|c|c|c||c|c|c|c|c|}\hline
Statistics    &\multicolumn{2}{c||}{Fermions} & \multicolumn{5}{c|}{Bosons } \\\hline
Fields      &
      $~~N~~$ & 
      $~~E~~$ & 
      $~~\Phi_1~~$ & 
      $~~\Phi_2~~$ & 
      $~~\varphi~~$ &
      $~~s^+_1~~$ &
      $~~s^+_3~~$ 
       \\ \hline
      $~~SU(2)_L~~$ & $\bf{1}$ & $\bf{1}$ & $\bf{2}$ & $\bf{2}$ & $\bf{1}$ & $\bf{1}$ & $\bf{1}$ \\ \hline
      $~~U(1)_Y~~$ & $0$ & $-1$ & $\frac12$ & $\frac12$ & $0$  & $1$  & $1$ \\ \hline
      $~~U(1)_{H}~~$ & $1$ & $-2$ & $1$ & $0$ & $-1$ & $1$ & $3$ \\ \hline
    \end{tabular}
  \end{center}
  \caption{Charge assignment for new vector-like lepton and scalar fields including SM-like Higgs $\Phi_2$. }
  \label{tab:charge}
\end{table}

We consider a model that is based on hidden $U(1)_H$ gauge symmetry. 
In the model we introduce vector-like $SU(2)_L$ singlet neutral and charged leptons $N$ and $E$, two Higgs doublets $\Phi_1$ and $\Phi_2$, singlet scalar $\varphi$, and 
$SU(2)_L$ singlet charged scalar fields $s_1^+$ and $s_3^+$. 
Here $\Phi_2$ is the SM-like Higgs.
The $U(1)_H$ charges of $\{N, E, \Phi_1, \varphi, s_1^+, s^+_3 \}$ are chosen as $\{1, -2, 1, -1, 1, 3 \}$ while other fields including the SM fermions are neutral under the hidden charge.
The charge assignment in the model is summarized in TABLE~\ref{tab:charge}.
In our scenario scalar fields $\{\Phi_1, \Phi_2, \varphi \}$ will develop vacuum expectation values(VEVs) to break $U(1)_H$ and electroweak symmetry.
The neutral fermion $N_{L(R)}$ plays a role of heavy neutrino in inverse seesaw mechanism where mass term $\overline{N^c_{L(R)}}N_{L(R)}$ is forbidden at tree level due to $U(1)_H$ symmetry. The charged contents $E$, $s^+_1$ and $s^+_3$ are introduced to realize inverse seesaw mechanism at one-loop level as we discuss below. In addition, these new particles contribute to muon $g-2$ without chiral suppression by the SM charged lepton mass. 

The Yukawa interactions of our model are given by 
\begin{align}
\mathcal{L}_{\rm Yukawa} = & 
y_\ell \overline{L_L} e_R \Phi_2 + y_\nu \overline{L_L} \tilde{\Phi}_1 N_R + y_1 \overline{N^c_R} E_R s_1^+ 
+M_N\overline{N_L} N_R
\nonumber \\
& + y_2 \overline{E_R} N_L s^-_3 + y_3 \overline{N^c_L} E_L s^+_1 + y_4 \overline{E_L} N_R s^-_3 + y_5 \overline{N_L} e_R s^{+}_1 + h.c. \ ,
\end{align}
where we omitted flavor indices and quark Yukawa sector, 
and define $\tilde{\Phi}_{1} = i \sigma_2 \Phi^*_{1}$ and $s^-_{1,3} = (s^+_{1,3})^*$.
Scalar potential is also written by
\begin{align}
V = & \mu_1^2 \Phi_1^\dagger \Phi_1 + \mu_2^2 \Phi_2^\dagger \Phi_2 + \mu_\varphi^2 \varphi^* \varphi + \mu_{s_1}^2 s^+_1 s^-_1 + \mu_{s_3}^2 s^+_3 s^-_3 + \mu_{12} (\Phi_2^\dagger \Phi_1 \varphi + h.c.) \nonumber \\
& + \mu_s (\Phi_1^T i \sigma_2 \Phi_2 s_1^- + h.c.) + \frac12 \lambda_1 (\Phi_1^\dagger \Phi_1)^2 + \frac12 \lambda_2 (\Phi_2^\dagger \Phi_2)^2 + \lambda_3 (\Phi_1^\dagger \Phi_1)(\Phi_2^\dagger \Phi_2)
\nonumber \\
& + \lambda_4 (\Phi_1^\dagger \Phi_2)(\Phi_2^\dagger \Phi_1) + \lambda_{s_1} (s^+_1 s^-_1)^2 + \lambda_{s_3} (s^+_3 s^-_3)^2 + \lambda_{\Phi_1 \varphi} (\Phi_1^\dagger \Phi_1)(\varphi^* \varphi) 
\nonumber \\
& + \lambda_{\Phi_2 \varphi} (\Phi_2^\dagger \Phi_2)(\varphi^* \varphi) + \lambda_{\Phi_1 s_1} (\Phi_1^\dagger \Phi_1)(s_1^+ s_1^-) + \lambda_{\Phi_1 s_3} (\Phi_1^\dagger \Phi_1)(s_3^+ s_3^-) 
\nonumber \\
& + \lambda_{\Phi_2 s_1} (\Phi_2^\dagger \Phi_2)(s_1^+ s_1^-) + \lambda_{\Phi_2 s_3} (\Phi_2^\dagger \Phi_2)(s_3^+ s_3^-) + \lambda_{s_1 s_3} (s_1^+ s_1^-)(s_3^+ s_3^-)
\nonumber \\
& + (\lambda_{ss} s_3^+ s_1^- \varphi \varphi + c.c.),
\end{align}
where coupling constants are taken to be real.

\subsection{Scalar sector}

In our scenario scalar fields $\{\Phi_1, \Phi_2, \varphi \}$ develop VEVs and they are written by
\begin{equation}
\Phi_i = \begin{pmatrix} w^+_i \\ \frac{1}{\sqrt{2}} (v_i + h_i + i z_i)  \end{pmatrix}, \quad \varphi = \frac{1}{\sqrt{2}} (v_\varphi + \phi^0 + i z'),
\end{equation}
where $v_i$ and $v_\varphi$ are the VEVs of $\Phi_i$ and $\varphi$.

For charged scalar sector, we rewrite $w^+_{1,2}$ by Higgs basis as 
\begin{equation}
\begin{pmatrix} w_1^+ \\ w_2^+ \end{pmatrix} =
\begin{pmatrix} \cos \beta & - \sin \beta \\ \sin \beta & \cos \beta \end{pmatrix}
\begin{pmatrix} w^+ \\ H^+ \end{pmatrix},
\end{equation}
where $\tan \beta = v_1/v_2$ and $w^+$ is identified as NG boson absorbed by $W^+$ boson.
Then we obtain mass terms for charged scalar fields as follows
\begin{align}
\mathcal{L_C} = & \ m^2_{H^\pm} H^+ H^- + \mu^2_{s_1} s_1^+ s_1^- + \mu^2_{s_3} s^+_3 s^-_3 \nonumber \\
& + m^2_{13} (s^+_3 s^-_1 + s^-_3 s^+_1) + m^2_{Hs_1} (H^+ s_1^- + H^- s^+_1),
\end{align}
where 
\begin{align}
m^2_{H^\pm} = \frac{\mu_{12} v_\varphi}{\sin \beta \cos \beta} - \frac{v^2 \lambda_4}{2}, \quad 
m^2_{13} = \frac{v^2_\varphi \lambda_{ss}}{2}, \quad m^2_{H s_1} = - \frac{v \mu_s}{\sqrt{2}}.
\end{align}
Thus we have 3 by 3 mass matrix in the basis of $\{H^+, s_1^+, s_3^+ \}$.
In our analysis, we require mixing between $s_1^\pm$ and $s_3^\pm$ is small by choosing parameter in the potential;
it is motivated to obtain small Majorana mass terms of $N_{L(R)}$ in realizing inverse seesaw mechanism.
In this case we can approximately obtain mass eigenstate $\{H_1^+, H_2^+, H_3^+ \}$ as 
\begin{align}
\begin{pmatrix} H^+ \\ s^+_1 \\ s^+_3 \end{pmatrix} & \simeq 
\begin{pmatrix} 1 & 0 & 0 \\ 0 & c_A & -s_A \\ 0 & s_A & c_A \end{pmatrix} 
\begin{pmatrix} c_B & - s_B & 0 \\ s_B & c_B & 0 \\ 0 & 0 & 1 \end{pmatrix}
\begin{pmatrix} H_1^+ \\ H_2^+ \\ H_3^+ \end{pmatrix},
\label{eq:scalar-masseigenstates}
\end{align}
where $c_{A[B]}(s_{A[B]}) = \cos A[B] (\sin A[B])$ with $A$ and $B$ being mixing angles ($A \ll B$ by our assumption).
The mass eigenvalues are denoted as $m_{H_i^\pm}$ for $H_i^\pm$.

We have three degrees of freedom for CP-even neutral scalar boson sector that is the same situation as two Higgs doublet with one singlet scalar case.
There is also one CP-odd neutral scalar boson that is similar to the case of two Higgs doublet model.
In this paper we do not discuss phenomenology related to neutral scalar bosons and we just assume mass spectrum and parameters in the potential 
satisfying current experimental/theoretical constraints.

\subsection{Gauge sector}

The gauge sector associated with $U(1)_Y$ and $U(1)_H$ is written by
\begin{equation}
\mathcal{L}_{\rm gauge} = - \frac14 B_{\mu \nu} B^{\mu \nu} - \frac14 B'_{\mu \nu} B'^{\mu \nu} - \frac12 \epsilon B_{\mu \nu} B'^{\mu \nu}
\end{equation} 
where $B_{\mu \nu}$ and $B'_{\mu \nu}$ are the gauge field strength tensors regarding $U(1)_Y$ and $U(1)_H$ respectively, and the last term is kinetic mixing one with kinetic mixing parameter $\epsilon$.
The kinetic terms can be diagonalized by the following transformation
\begin{equation}
\left(\begin{array}{c}
\tilde{B}^\prime_\mu\\
\tilde{B}_\mu\\
\end{array}\right)=\left(\begin{array}{cc}
\sqrt{1 - \epsilon^2} & 0 \\
\epsilon & 1 \\
\end{array}\right)\left(\begin{array}{c}
B^\prime_\mu\\
B_\mu\\
\end{array}\right).
\label{eq:kinetic}
\end{equation}
%
The gauge boson masses are generated from kinetic terms of VEV developing scalar fields 
\begin{eqnarray}
\mathcal{L}_{\text{kin}} = (D_{\mu}H_1)^\dagger (D^{\mu} H_1) + (D_{\mu}H_2)^\dagger (D^{\mu} H_2) + (D_{\mu}\varphi)^\dagger (D^{\mu} \varphi).
\label{eq:scalar-kinetic}
\end{eqnarray}
We write the covariant derivatives as follows
\begin{eqnarray}
D_{\mu}H_1 &=& \Big( \partial_\mu + i g_2 \frac{\tau^a}{2} W_{\mu}^a +  i\frac{g_1}{2}\tilde{B}_\mu + \Big( i \frac{g_1}{2} \rho -i g_X \frac{\rho}{\epsilon} \Big)\tilde{B}^\prime_\mu \Big)H_1, \nonumber \\
D_{\mu}H_2 &=& \Big(\partial_\mu + ig_2 \frac{\tau^a}{2}W_{\mu}^a + i\frac{g_1}{2}\tilde{B}_\mu + i\frac{g_1}{2}\rho \tilde{B}^\prime_\mu  \Big)H_2, \nonumber \\
D_{\mu}\varphi &=& \Big( \partial_\mu + ig_X  \frac{\rho}{\epsilon}\tilde{B}^\prime_\mu \Big)\varphi, 
\label{Covariant derivatives}
\end{eqnarray}
where we parameterize $\rho=-\frac{\epsilon}{\sqrt{1-\epsilon^2}}$,
$g_2$, $g_1$ and $g_X$ are respectively gauge couplings of $SU(2)_L$, $U(1)_Y$ and $U(1)_H$, $W^a$ is the $SU(2)_L$ gauge field, and $\tau^a$ is the Pauli matrix. 
After symmetry breaking, we obtain $W$ boson mass as in the SM and neutral gauge boson mass terms such that 
\begin{align}
& \mathcal{L}_{ZZ'} = \frac{1}{2} m_{Z_{\rm SM}}^2 \tilde Z_\mu \tilde Z^\mu + \Delta M^2 \tilde Z_\mu \tilde B'^\mu + \frac12 m_{B'}^2 \tilde B'_\mu \tilde B'^\mu, \\
& m_{Z_{\rm SM}}^2 = \frac14 (g_1^2 + g_2^2) v^2, \quad m_{B'}^2 =  \left( \frac12 g_1 \rho - g_X \frac{\rho}{\epsilon} \right)^2 v_1^2 + \frac14 g_1^2 \rho^2 v_2^2 + \frac12 g_X^2 \frac{\rho^2}{\epsilon^2} v_\varphi^2, \nonumber \\
& \Delta M^2 = \frac{1}{2} \sqrt{g_1^2 +g_2^2} \left( \frac12 g_1 \rho - g_X \frac{\rho}{\epsilon} \right) v_1^2 + \frac{1}{4} \sqrt{g_1^2 + g_2^2} g_1 \rho v_2^2,
\end{align}
where $\tilde Z_\mu \equiv - \cos \theta_W W^3_\mu + \sin \theta_W \tilde B_\mu$ with $\theta_W$ being the Weinberg angle.
By diagonalizing the mass terms the physical masses of the neutral gauge bosons are obtained as
\begin{eqnarray}
m_Z^2 &=& \frac{1}{2}\Big[M_{Z,SM}^2 + M_Z^{\prime 2} + \sqrt{(M_{Z,SM}^2 - M_{Z^\prime}^2)^2 + 4\Delta^4} \Big] \nonumber \\
m_{Z^{\prime}}^2 &=& \frac{1}{2}\Big[M_{Z,SM}^2 + M_Z^{\prime 2} - \sqrt{(M_{Z,SM}^2 - M_{Z^\prime}^2)^2 + 4\Delta^4} \Big].
\end{eqnarray}
To avoid constraint from the $\rho$-parameter, we require $\epsilon \ll 1$ and $m_Z^2 \ll g_X^2 v_\varphi^2$ so that SM $Z$ boson mass is $m_Z\simeq m_{Z_{\rm SM}}$ and $m_{Z'} \simeq m_{B'}$.
The rotation matrix diagonalizing neutral gauge boson sector is given by
\begin{align}
\left(\begin{array}{c}
\tilde{Z}_\mu\\
\tilde{B}^\prime_\mu\\
\end{array}\right) &=
\left(\begin{array}{cc}
\cos\theta & \sin\theta\\
-\sin\theta & \cos\theta \\
\end{array}\right)
\left(\begin{array}{c}
Z_\mu\\
Z^\prime_\mu\\
\end{array}\right) \\
\tan 2\theta &= \frac{2\Delta^2}{M_{Z,SM}^2 - M_{Z^\prime}^2}
\end{align}
where $Z_\mu$ and $Z^\prime_\mu$ are the physical gauge bosons that are the SM $Z$ boson and extra $Z'$ gauge boson.
For neutrino mass and flavor physics $Z'$ boson does not contribute and we just assume it is heavy enough and $Z$-$Z'$ mixing is small 
to avoid phenomenological constraints.

\subsection{Neutrino mass}

\begin{figure}[tb]
\includegraphics[width=10cm]{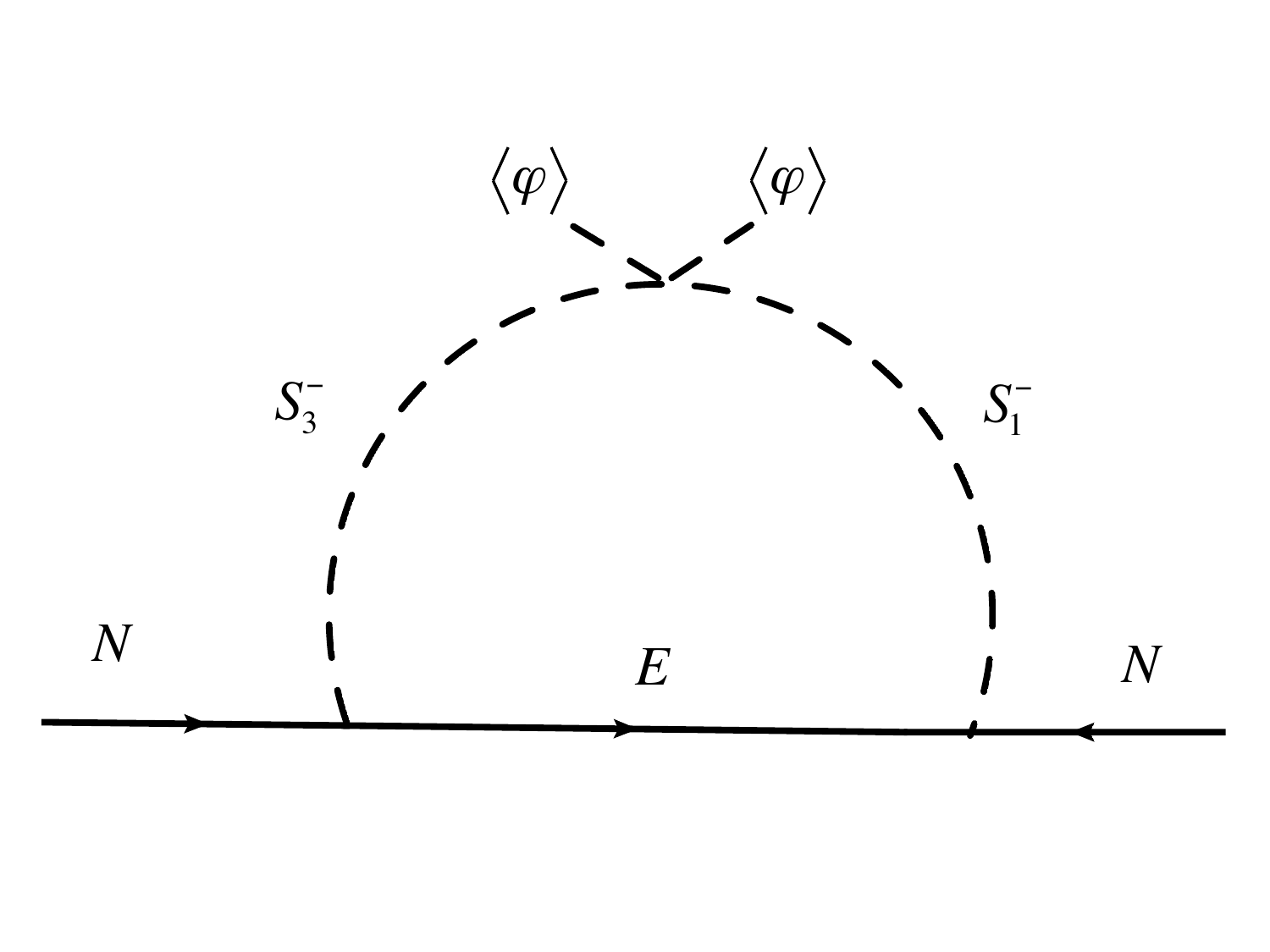} \ 
\caption{One-loop diagram that generates Majorana mass term of $N_{L(R)}$.}
\label{fig:diagram1}
\end{figure}

After scalar fields developing VEVs mass terms of neutral fermions at tree level are written by
\begin{align}
\mathcal{L}_M = M_N \overline{N_L} N_R + m_D \overline{\nu_L} N_R + h.c., 
\end{align}
where $m_D \equiv y_\nu v_{1}/\sqrt{2}$.
Thus structure of tree level mass matrix is
\begin{equation}
M_\nu = \left( \begin{array}{ccc} 0 & m_D^* & 0 \\ m_D^T & 0 & M_{N}^\dag \\ 0 & M_{N}^* & 0 \end{array} \right)
\label{m-nu-1}
\end{equation}
in the $(\nu_L~N_R^c~N_L)$ basis. 
The 22 and 33 block components of $M_\nu$ can be generated at loop level.
The relevant interactions to generate them are 
\begin{align}
\mathcal{L} \supset & \ y_1 \overline{N_R^c} E_R (c_A s_B H_1^+ + c_A c_B H_2^+ - s_A H_3^+) + y_2 \overline{E_R} N_L (s_A s_B H_1^- + s_A c_B H_2^- + c_A H_3^-)  \nonumber \\
& + y_3 \overline{N^c_L} E_L (c_A s_B H_1^+ + c_A c_B H_2^+ - s_A H_3^+) +  y_4 \overline{E_L} N_R (s_A s_B H_1^- + s_A c_B H_2^- + c_A H_3^-) \nonumber \\ 
& + h. c.  \, ,
\end{align}
where we used Eq.~\eqref{eq:scalar-masseigenstates} to write mass eigenstates for charged scalars.
The 22 and 33 elements of $M_\nu$ are generated by the diagram in Fig.~\ref{fig:diagram1}. 
We then calculate one-loop diagrams and find 
\begin{align}
(\delta m_{22})_{ij} &= \frac{s_A c_A}{(4 \pi)^2} \sum_{\alpha = 1,2,3}  (y_{4}^\dagger)_{j \alpha }  (y_{1}^\dagger)_{ \alpha i} M_{E_\alpha}  I (M^2_{E_\alpha}, m^2_{H^+_{1,2,3}}), \label{eq:m22}
\\
(\delta m_{33})_{ij} &= \frac{s_A c_A}{(4 \pi)^2} \sum_{\alpha = 1,2,3}  (y_{3})_{i \alpha} (y_{2})_{\alpha j} M_{E_\alpha} I (M^2_{E_\alpha}, m^2_{H^+_{1,2,3}}),
\label{eq:m33}
\end{align}
where $I(M^2_{E_\alpha}, m^2_{H^+_a})$ is function of extra charged lepton/Higgs masses from loop integration.
The explicit form of $I(M^2_{E_\alpha}, m^2_{H^+_{1,2,3}})$ is given by
\begin{align}
I(M^2_{E_\alpha}, m^2_{H^+_{1,2,3}}) & = \int [dX]_{ 4} \left[ 2 \frac{s^2_B m^2_{H_1^+} + c^2_B m^2_{H_2^+} - m^2_{H_3^+}}{M^2_{E_\alpha} x + m^2_{H_1^+} y+ m^2_{H_2^+} z + m^2_{H_3^+} w} \right. \nonumber \\
& \left. \qquad \qquad  \qquad +  \frac{s^2_B m^2_{H_2^+}m^2_{H_3^+} + c^2_B m^2_{H_1^+} m^2_{H_3^+} - m^2_{H_1^+} m^2_{H_2^+}}{(M^2_{E_\alpha} x + m^2_{H_1^+} y+ m^2_{H_2^+} z + m^2_{H_3^+} w)^2} \right],
\end{align}
where $\int [dX]_{ 4} = \int_0^1 dx dy dz dw \delta(1-x-y-z-w)$.

The loop induced elements $\delta m_{22}$ and $\delta m_{33}$ are naturally taken to be small and we assume $\delta m_{22(33)} \ll m_D \ll M_N$.
Then we find the active neutrino mass matrix via IS mechanism such as 
\begin{equation}
m_\nu \simeq (m_D M^{-1}_N) \delta m_{33} (m_D M^{-1}_N)^T.
\end{equation}
Note that $\delta m_{22}$ element does not contribute to the active neutrino mass matrix at the leading order.
To simplify our analysis we factor out scale of mass scale in our new physics sector such that
\begin{equation}
\{ M_{N_\alpha}, M_{E_\alpha}, m_{H^\pm_i} \} = \mu_{\rm NP} \times \{ \tilde M_{N_\alpha}, \tilde M_{E_\alpha}, \tilde m_{H^\pm_i}\},
\end{equation}
where $\mu_{\rm NP}$ has mass scale and parameters in RHS with "$\sim$" are dimensionless ones representing the ratio of mass parameters. 
We then write active neutrino mass matrix by
\begin{align}
\label{eq:mnu}
m_\nu  & \simeq \frac{v^2 s_\beta^2 }{2 \mu_{\rm NP}} (y_\nu \tilde{M}^{-1}_N) \delta \tilde{m}_{33} (y_\nu \tilde{M}^{-1}_N)^T \nonumber \\
& \equiv \kappa (y_\nu \tilde{M}^{-1}_N) \delta \tilde{m}_{33} (y_\nu \tilde{M}^{-1}_N)^T \nonumber \\
&  \equiv \kappa \tilde{m}_\nu, 
\end{align}
where $\delta m_{33} = \mu_{\rm NP} \delta \tilde{m}_{33}$.
The neutrino mass matrix $m_\nu$ is diagonalized by a unitary matrix $V_{\nu}$ by $D_\nu=|\kappa| \tilde D_\nu= V_{\nu}^T m_\nu V_{\nu}=|\kappa| V_{\nu}^T \tilde m_\nu V_{\nu}$.
We then determine $|\kappa|$ using squared mass differences observed in neutrino oscillation experiments such that
\begin{align}
(\mathrm{NO}):\  |\kappa|^2= \frac{|\Delta m_{\rm atm}^2|}{\tilde D_{\nu_3}^2-\tilde D_{\nu_1}^2},
\quad
(\mathrm{IO}):\  |\kappa|^2= \frac{|\Delta m_{\rm atm}^2|}{\tilde D_{\nu_2}^2-\tilde D_{\nu_3}^2},
 \end{align}
where $\Delta m_{\rm atm}^2$ corresponds to the atmospheric neutrino mass-squared splitting, and NO and IO respectively indicate the normal and the inverted ordering of neutrino mass. 
Using the determined $|\kappa|$, the solar mass squared splitting can be given by
\begin{align}
\Delta m_{\rm sol}^2=  |\kappa|^2 ({\tilde D_{\nu_2}^2-\tilde D_{\nu_1}^2}),
 \end{align}
 which is compared with observed value in numerical analysis.
 %
The observed neutrino mixing matrix is defined by $U= V_\nu$~\cite{Maki:1962mu}
since we assume the charged-lepton mass matrix to be diagonal, where
elements of the matrix are parametrized by three mixing angles $\theta_{ij} (i,j=1,2,3; i < j)$, one CP violating Dirac phase $\delta_{CP}$,
and two Majorana phases $\alpha_{21},\alpha_{31}$.
The mixing matrix is thus explicitly written by 
\begin{equation}
U = 
\begin{pmatrix} c_{12} c_{13} & s_{12} c_{13} & s_{13} e^{-i \delta_{CP}} \\ 
-s_{12} c_{23} - c_{12} s_{23} s_{13} e^{i \delta_{CP}} & c_{12} c_{23} - s_{12} s_{23} s_{13} e^{i \delta_{CP}} & s_{23} c_{13} \\
s_{12} s_{23} - c_{12} c_{23} s_{13} e^{i \delta_{CP}} & -c_{12} s_{23} - s_{12} c_{23} s_{13} e^{i \delta_{CP}} & c_{23} c_{13} 
\end{pmatrix}
\begin{pmatrix} 1 & 0 & 0 \\ 0 & e^{i \frac{\alpha_{21}}{2}} & 0 \\ 0 & 0 &  e^{i \frac{\alpha_{31}}{2}} \end{pmatrix}.
\end{equation}
where $c_{ij}$ and $s_{ij}$ stand for $\cos \theta_{ij}$ and $\sin \theta_{ij}$ ($i,j=1-3$), respectively. 
Then, each of the mixing angle is obtained in terms of the component of $U$ such that
\begin{align}
\sin^2\theta_{13}=|U_{e3}|^2,\quad 
\sin^2\theta_{23}=\frac{|U_{\mu3}|^2}{1-|U_{e3}|^2},\quad 
\sin^2\theta_{12}=\frac{|U_{e2}|^2}{1-|U_{e3}|^2}.
\end{align}
The Dirac phase  $\delta_{CP}$ is also given by computing  the Jarlskog invariant as
\begin{align}
\sin \delta_{CP} &= \frac{\text{Im} [U_{e1} U_{\mu 2} U_{e 2}^* U_{\mu 1}^*] }{s_{23} c_{23} s_{12} c_{12} s_{13} c^2_{13}} ,\quad
\cos \delta_{CP} = -\frac{|U_{\tau1}|^2 -s^2_{12}s^2_{23}-c^2_{12}c^2_{23}s^2_{13}}{2 c_{12} s_{12} c_{23} s_{23}s_{13}} ,
\end{align}
where $\delta_{CP}$ be subtracted from $\pi$ when $\cos \delta_{CP}$ is negative.
Also Majorana phase $\alpha_{21},\ \alpha_{31}$ are found as follows
\begin{align}
&
\sin \left( \frac{\alpha_{21}}{2} \right) = \frac{\text{Im}[U^*_{e1} U_{e2}] }{ c_{12} s_{12} c_{13}^2} ,\quad
  \cos \left( \frac{\alpha_{21}}{2} \right)= \frac{\text{Re}[U^*_{e1} U_{e2}] }{ c_{12} s_{12} c_{13}^2}, \
%
,\\
&
 \sin \left(\frac{\alpha_{31}}{2}  - \delta_{CP} \right)=\frac{\text{Im}[U^*_{e1} U_{e3}] }{c_{12} s_{13} c_{13}},
\quad 
 \cos \left(\frac{\alpha_{31}}{2}  - \delta_{CP} \right)=\frac{\text{Re}[U^*_{e1} U_{e3}] }{c_{12} s_{13} c_{13}},
\end{align}
where $\alpha_{21}/2,\ \alpha_{31}/2-\delta_{CP}$
are subtracted from $\pi$ if $ \cos \left( \frac{\alpha_{21}}{2} \right),\  \cos \left(\frac{\alpha_{31}}{2}  - \delta_{CP} \right)$ are negative.
Finally the effective mass for the neutrinoless double beta decay is written by
\begin{align}
\langle m_{ee}\rangle=|\kappa||\tilde D_{\nu_1} \cos^2\theta_{12} \cos^2\theta_{13}+\tilde D_{\nu_2} \sin^2\theta_{12} \cos^2\theta_{13}e^{i\alpha_{21}}+\tilde D_{\nu_3} \sin^2\theta_{13}e^{-2i\delta_{CP}}|,
\end{align}
where its observed value could be measured by KamLAND-Zen in future~\cite{KamLAND-Zen:2016pfg}. 

In addition to the neutrino mass data, we need to consider a constraint from non-unitarity that should be taken into account for IS model. 
Here the non-unitarity matrix is denoted by $U'$ and it represents the deviation from the unitarity parametrized by the following form 
\begin{align}
U'\equiv \left(1-\frac12 FF^\dag\right) U,
\end{align}
where $F\equiv  (M^*_{N})^{-1} m^T_D$ is a hermitian matrix given by elements of neutral fermion matrix $M_\nu$.
The global constraints are found combining several experimental results such as the the effective Weinberg angle, SM $W$ boson mass, several ratios of $Z$ boson fermionic decays, invisible decay of $Z$, electroweak universality, measurements of Cabbibo-Kobayashi-Maskawa matrix, and lepton flavor violations~\cite{Fernandez-Martinez:2016lgt}.
These results provide the following constraints for $|FF^\dag|$~\cite{Agostinho:2017wfs, Das:2017ski}
\begin{align}
|FF^\dag|\le  
\left[\begin{array}{ccc} 
2.5\times 10^{-3} & 2.4\times 10^{-5}  & 2.7\times 10^{-3}  \\
2.4\times 10^{-5}  & 4.0\times 10^{-4}  & 1.2\times 10^{-3}  \\
2.7\times 10^{-3}  & 1.2\times 10^{-3}  & 5.6\times 10^{-3} \\
 \end{array}\right]. \label{eq:const-non-unitarity}
\end{align} 
In numerical analyses below we impose the above constraints.

\subsection{Lepton flavor violation and muon $g-2$}

 \begin{figure}[tb]
\includegraphics[width=10cm]{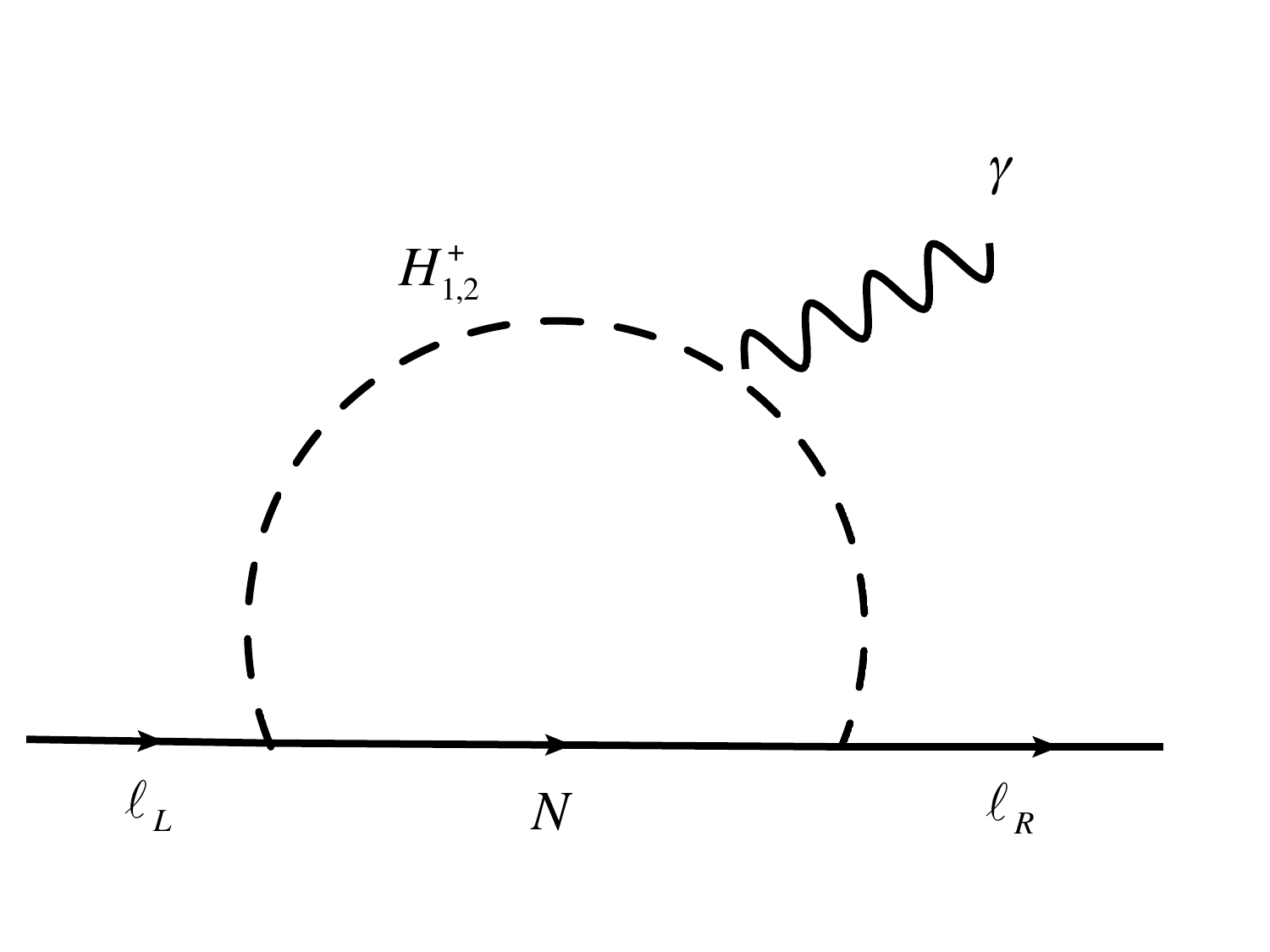} \ 
\caption{One-loop diagram contributing lepton $g-2$, EDM and LFV decay amplitude.}
\label{fig:diagram2}
\end{figure}

In the model muon $g-2$ and LFV decay amplitudes can be induced at one loop level without chiral suppression.
The relevant Lagrangian terms are given by
\begin{align}
\label{eq:YukawaLFV}
\mathcal{L}_{LFV} = - y_\nu s_\beta \overline{\ell_L} N_R (c_B H_1^- - s_B H_2^-) + y_5 \overline{N_L} e_R (c_A s_B H_1^+ + c_A c_B H_2^+ - s_A H^+_3) +h.c. \, .
\end{align}
The dominant one loop diagram is given in Fig.~\ref{fig:diagram2}.
Calculating one loop diagram, we obtain new physics contribution to electron and muon $g-2$ as 
\begin{equation}
\Delta a_{e, \mu} = - m_\mu  {\rm Re} [(a_R + a_L)_{11, 22}],
\end{equation}
where the amplitudes are given by
\begin{align}
(a_{R(L)})_{ij} & = (a_{R_{1(2)}} )_{ij} + (a_{L_{1(2)}} )_{ij}, \nonumber \\
(a_{R_{1(2)}})_{ij} & \simeq \frac{y^{jk}_\nu y^{ki}_5}{16 \pi^2} s_\beta c_A c_B s_B M_{N_k} \int_0^1 dx \frac{x}{x M^2_{N_k} + (1-x) m^2_{H_{1(2) }^+ }} , \nonumber \\ 
(a_{L_{1(2)}})_{ij} & \simeq \frac{(y_\nu^\dagger)^{ki} (y^{\dagger}_5)^{jk} }{16 \pi^2} s_\beta c_A c_B s_B M_{N_k} \int_0^1 dx \frac{x}{x M^2_{N_k} + (1-x) m^2_{H_{1(2) }^+ }} ,
\end{align}
where we ignored charged lepton masses and $N_k$ is treated as Dirac fermion approximately.
 Thus $\Delta a_{e,\mu}$ depends on real part of the sum of products of Yukawa couplings. For example, if all the $M_{N_k}$ value are degenerate we have $|\Delta a_{e(\mu)}| \propto |y_\nu^{1(2)k}y_5^{k1(2)}| \cos({\rm Arg}[y_\nu^{1(2)k}y_5^{k1(2)}]) + |(y_\nu^\dagger)^{k1(2)}(y^\dagger_5)^{1(2)k}| \cos({\rm Arg} [ (y_\nu^\dagger)^{k1(2)}(y^\dagger_5)^{1(2)k}] )$.  

For electron $g-2$, the SM prediction depends on the value of the fine-structure constant $\alpha_{em}$ where 
the latest measurements based on Cesium and Rubidium show different values~\cite{Parker:2018vye, Morel:2020dww}.
We then have difference between the SM prediction and observed value of electron $g-2$ such that 
\begin{align}
& \Delta a_e ({\rm Cesium}) = (-8.8 \pm 3.6) \times 10^{-13}, \label{eq:ae-1} \\
& \Delta a_e ({\rm Rubidium}) = (4.8 \pm 3.0) \times 10^{-13} \label{eq:ae-2}.  
\end{align}
In numerical analysis below,  our theoretical value of $\Delta a_e$ is compared with these observations.

In addition to electron and muon $g-2$, we have contribution to EDM from the same loop diagram.
Electron and muon EDMs are given by
\begin{equation}
d_{e, \mu} = \frac{e}{2} {\rm Im}[(a_R - a_L)_{11, 22}],
\end{equation} 
where $e$ is the electromagnetic coupling.
In contrast to magnetic moment $d_{e,\mu}$ depends on imaginary part of the sum of products of Yukawa couplings. For example, if all the $M_{N_k}$ value are degenerate we have $| d_{e(\mu)}| \propto |y_\nu^{1(2)k}y_5^{k1(2)}| \sin({\rm Arg}[y_\nu^{1(2)k}y_5^{k1(2)}]) - |(y_\nu^\dagger)^{k1(2)}(y^\dagger_5)^{1(2)k}| \sin({\rm Arg} [ (y_\nu^\dagger)^{k1(2)}(y^\dagger_5)^{1(2)k}] )$.  
We adopt the latest experimental constraints on $|d_{e, \mu}|$~\cite{Roussy:2022cmp, Muong-2:2008ebm};
\begin{equation}
|d_e| < 6.3 \times 10^{-17} {\rm GeV}^{-1}, \quad |d_\mu| < 2.76 \times 10^{-6} {\rm GeV}^{-1}.
\end{equation}
Note that we have tendency that if $g-2$ is larger EDM is also larger. For simplicity, we shall take only one phase for Yukawa couplings as $y_\nu^{ij} = |y_\nu^{ij}| e^{i\phi}$ and $y_5^{ij} = |y_5^{ij}| e^{i\phi}$ (and degenerate $M_{N_k}$). Then we get $|\Delta a_e| \propto  (|y_\nu^{1k} y_5^{k1} | + |y_\nu^{k1} y_5^{1k} |) |\cos 2\phi |$ and $|d_e| \propto  (|y_\nu^{1k} y_5^{k1} | + |y_\nu^{k1} y_5^{1k} |) | \sin 2\phi |$. Since phase is arbitrary number we expect larger value of  $|d_\ell |$  for larger $|\Delta a_\ell |$ unless we tune phases to make EDM small.

The same one-loop diagram also induces LFV decay processes of charged leptons $\ell_i \to \ell_j \gamma$.
The decay BRs of the processes are given by
\begin{equation}
BR (\ell_i \to \ell_j \gamma) = \frac{48 \pi^3 C_{ij} \alpha_{\rm em}}{G_F^2 m_i^2} m_i^3 (|(a_R)_{ij}|^2+|(a_L)_{ij}|^2),
\end{equation}
where $(C_{21}, C_{31}, C_{32}) = (1, 0.1784, 0.1736)$.
The current experimental upper limits for these BRs are given 
by~\cite{MEGII:2023ltw, MEG:2016leq, BaBar:2009hkt,Renga:2018fpd}
  \begin{align}
  B(\mu\rightarrow e\gamma) &\leq  3.1\times10^{-13}, \quad 
  B(\tau\rightarrow \mu\gamma)\leq4.4\times10^{-8}, \quad  
  B(\tau\rightarrow e\gamma) \leq3.3\times10^{-8}~.
 \label{expLFV}
 \end{align}
We impose these constraints in numerical analysis.

\subsection{One-loop modification in leptonic $Z$ boson decays}

 \begin{figure}[tb]
\includegraphics[width=10cm]{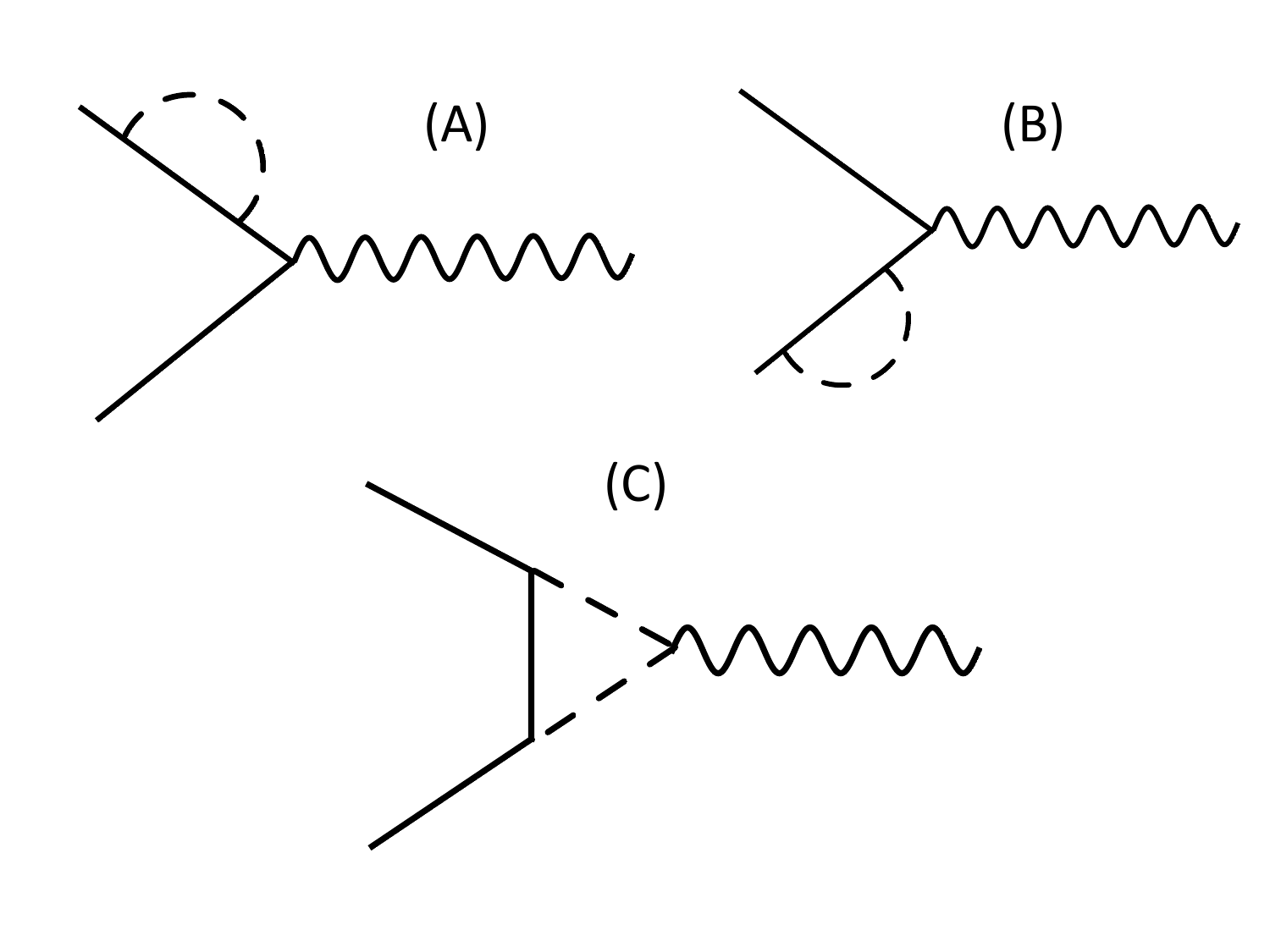} \ 
\caption{Diagrams for modifying interactions between $Z$ and SM charged leptons where dashed line indicate charged scalar bosons.}
\label{fig:diagram3}
\end{figure}

We consider the one-loop contributions to $Z$ boson decay into two charged leptons that are induced by the Yukawa interactions in Eq.~\eqref{eq:YukawaLFV}.
The relevant diagrams are shown in Fig.~\ref{fig:diagram3} where  the dashed line includes three charged scalar bosons and we consider all the possible combinations of Yukawa couplings in Eq.~\eqref{eq:YukawaLFV} for the diagrams.  
By these diagrams, interactions between $Z$ boson and charged leptons are modified as 
\begin{equation}
\mathcal{L}^{\rm SM + NP}_{Z \ell \bar \ell} = \frac{g_2}{c_W} \overline{\ell_i} \gamma^\mu \left[ \left(- \frac12 + s_W^2 \right) \left( \delta_{ij} + (\delta^{\rm NP}_{L})_{ij} \right) P_L
+ s^2_W  \left( \delta_{ij} + (\delta^{\rm NP}_{R})_{ij} \right) P_R \right] \ell_j
\end{equation}
where $g_2$ is the $SU(2)_L$ gauge coupling constant.
The new physics contribution is written by
\begin{equation}
(\delta^{\rm NP}_{R(L)})_{ij} 
=
 (\delta A_{R(L)})_{ij} + (\delta B_{R(L)})_{ij} +  (\delta C_{R(L)})_{ij} + (\delta D_{R(L)}),  
\end{equation}
where $\delta A_{R(L)}$, $\delta B_{R(L)}$, $\delta C_{R(L)}$ and $\delta D_{R(L)}$ are obtained by calculating loop diagrams in Fig.~\ref{fig:diagram3}.
Here the contribution $\delta A_{R(L)}$ and $\delta B_{R(L)}$ are obtained from diagram (C) in the figure with coupling combination of $y_\nu$ and $y_5$ for loop.
The contribution $\delta C_{R(L)}$ is obtained from summation of diagrams (A) and (B) with coupling combination of $y_\nu (y_\nu^\dagger)$ and $y_5 (y_5^\dagger)$ in the loops.
Also the contribution $\delta D_{R(L)}$ is obtained from summation of diagrams (A), (B) and (C) with coupling combination of $y_\nu (y_5)$ and $y_\nu^\dagger (y_5^\dagger)$ for loops. 
In our parameter region of interest, $\delta D_{R(L)}$ is dominant one and the approximated expression for dominant contribution is given by 
\begin{align}
(\delta D_R)_{ij}  & \simeq -  \frac{(y_5^\dagger)_{i \alpha} (y_5)_{\alpha j}}{(4\pi)^2} \int [dX]_3 \frac{z (1-z+x) m_Z^2}{x M^2_{E_\alpha} +(1-x) m^2_{H_2^+}}, \\
(\delta D_L)_{ij} & \simeq -  \frac{(y_\nu)_{i \alpha} (y_\nu^\dagger)_{\alpha j}}{(4\pi)^2} \int [dX]_3 \frac{z (1-z+x) m_Z^2}{x M^2_{E_\alpha} +(1-x) m^2_{H_1^+}},
\end{align}
where  $\int [dX]_{ 3} = \int_0^1 dx dy dz \delta(1-x-y-z)$ and we approximated the formula under $m_i^2 \ll \{M_{E_\alpha}^2, m^2_{H_1^\pm}, m^2_{H_2^\pm} \}$.
The full expressions of subdominant contributions $\delta A_{R(L)}$, $\delta B_{R(L)}$ and $\delta C_{R(L)}$ are summarized in the Appendix.
The widths of $Z$ boson into charged leptons are modified by new physics contribution such that
\begin{equation}
\Gamma^{\rm SM+NP}_{Z \to \ell_i \overline{\ell_j} } = \frac{g_2^2}{24 \pi c_W^2} m_Z \left[ \left( - \frac12 + s^2_W \right) \left| \delta_{ij} + (\delta_L^{NP} )_{ij} \right|^2 
+ S_W^4 \left|\delta_{ij} + (\delta_R^{NP} )_{ij} \right|^2 \right].
\end{equation}
Note that we have flavor non-conserving $Z$ boson decays from new physics contributions. Then we impose the constraints~\cite{ParticleDataGroup:2020ssz}:
\begin{align}
& {\rm BR}(Z \to e^\pm \mu^\mp) \leq 7.5 \times 10^{-7}, \quad {\rm BR}(Z \to e^\pm \tau^\mp) \leq 9.8 \times 10^{-6}, \nonumber \\
& {\rm BR}(Z \to \mu^\pm \tau^\mp) \leq 1.2 \times 10^{-5}.
\end{align}
For flavor conserving modes, we define deviation from the SM prediction by
\begin{equation}
\Delta {\rm BR} (Z \to \ell_i \overline{\ell_i})  = \frac{\Gamma^{\rm SM+NP}(Z \to \ell_i \overline{\ell_i} ) - \Gamma^{\rm SM}(Z \to \ell_i \overline{\ell_i} ) }{\Gamma_Z^{\rm tot}},
\end{equation}
where $\Gamma_Z^{\rm tot} = 2.4952 \pm 0.0023$ GeV is the total decay width of $Z$ boson. 
We impose the current bounds on the deviation for the lepton flavor conserving $Z$ boson decays which are given by~\cite{ParticleDataGroup:2020ssz}
\begin{align}
& |\Delta {\rm BR}(Z \to e^+ e^-)| \leq 4.2 \times 10^{-5}, \quad |\Delta {\rm BR}(Z \to \mu^+ \mu^-)| \leq 6.6 \times 10^{-5}, \nonumber \\
& |\Delta {\rm BR}(Z \to \tau^+ \tau^-)| \leq 1.2 \times 10^{-5} .
\end{align}
Notice that we do not consider $Z \to \nu \bar \nu$ decay mode since new physics effect is smaller and the experimental constraints are also less significant.

\section{Numerical analysis and phenomenology}

In this section we perform numerical analysis searching for allowed parameter points that satisfy neutrino data and  constraints from non-unitarity and LFVs.
Firstly we fix the ratio of mass parameters such that 
\begin{align}
& \{\tilde{m}_{H^\pm_1}, \tilde{m}_{H^\pm_2}, \tilde{m}_{H^\pm_3}, \tilde{M}_{E_1}, \tilde{M}_{E_2}, \tilde{M}_{E_3}, \tilde{M}_{N_1}, \tilde{M}_{N_2}, \tilde{M}_{N_3} \} \nonumber \\
&= \{1.0, \ 1.5, \ 2.0, \ 1.2, \ 1.7, \ 2.2, \ 1.1, \ 1.6, \ 2.1 \}.  \label{eq:masses}
\end{align}
Notice that neutrino mass matrix and LFV amplitudes are not much sensitive to the mass parameters and other choices do not change tendency of the results unless 
there is large hierarchy among them.
We also fix $s_\beta = 1/\sqrt{2}$, $s_{A(B)} =0.01(0.1)$, and $\mu_{NP}$ is determined by $|\kappa|$.
Then, as relevant parameters, we adopt $\{(y_\nu)_{ij}, (y_5)_{ij},  (\delta \tilde{m}_{33})_{ij}  \}$ where  we consider $(\delta \tilde{m}_{33})_{ij}$ as free parameter 
although it is given by other parameters as Eq.~\eqref{eq:m33} for simplicity.
Then, free parameters are globally scanned in the following region:
\begin{align}
& |(y_{\nu})_{ij}| \in [10^{-5}, 1.0], \quad |(y_5)_{ij}| \in [10^{-5}, 1.0], \nonumber \\
& |(\delta \tilde{m}_{33})_{ij}| \in \left[10^{-10}, \frac{s_A c_A}{(4 \pi)^2} {\rm Max}[ M_{E_\alpha} I (M^2_{E_\alpha}, m^2_{H^+_{1,2,3}})]  \right], \label{eq:scan}
\end{align}
where the upper bound of $|(\delta \tilde{m}_{33})_{ij}|$ is fixed based on Eq.~\eqref{eq:m33}.

For neutrino sector, we require $\{s_{12}, s_{13}, s_{23}, \Delta m^2_{\rm sol} \}$ satisfy $3 \sigma$ range of observed value in NuFIT 5.2~\cite{Esteban:2020cvm} where observed value of $\Delta m^2_{\rm atm}$ is used as input parameter 
choosing within 3$\sigma$ range randomly. 
We then search for allowed parameter sets of $(y_\nu)_{ij}$ satisfying the neutrino observation conditions and imposing the constraints in Eq.~\eqref{eq:const-non-unitarity}.
The new physics scale $\mu_{NP}$ is also obtained as output from relation in Eq.~\eqref{eq:mnu}.

Performing numerical analysis for neutrino sector, we obtain allowed values of Yukawa coupling $(y_\nu)_{ij}$ and $\mu_{NP}$. 
Here we do not show observables related to neutrinos since any pattern is possible by changing $\delta \tilde m_{33}$ and there is no prediction.
The neutrino data is used to constrain structure of  $(y_\nu)_{ij}$ and possible value of $\mu_{NP}$.
Then we calculate lepton $g-2$, deviations in $Z$ boson decays, EDM and LFVs using the allowed Yukawa coupling $(y_\nu)_{ij}$ and $\mu_{NP}$ with $(y_5)_{ij}$ scanned within the range in Eq.~\eqref{eq:scan} 
and search for the parameters providing sizable muon $g-2$ satisfying the other constraints.
It is found that deviations in $Z$ boson decays are much smaller than current constraints in the model and we do not show the plot for them.

 \begin{figure}[tb]
\includegraphics[width=7cm]{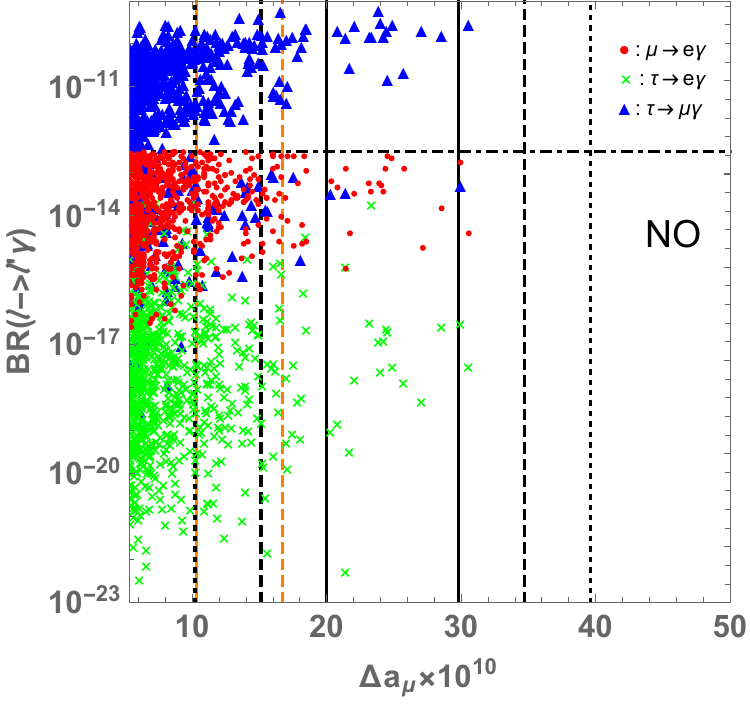} \ 
\includegraphics[width=7cm]{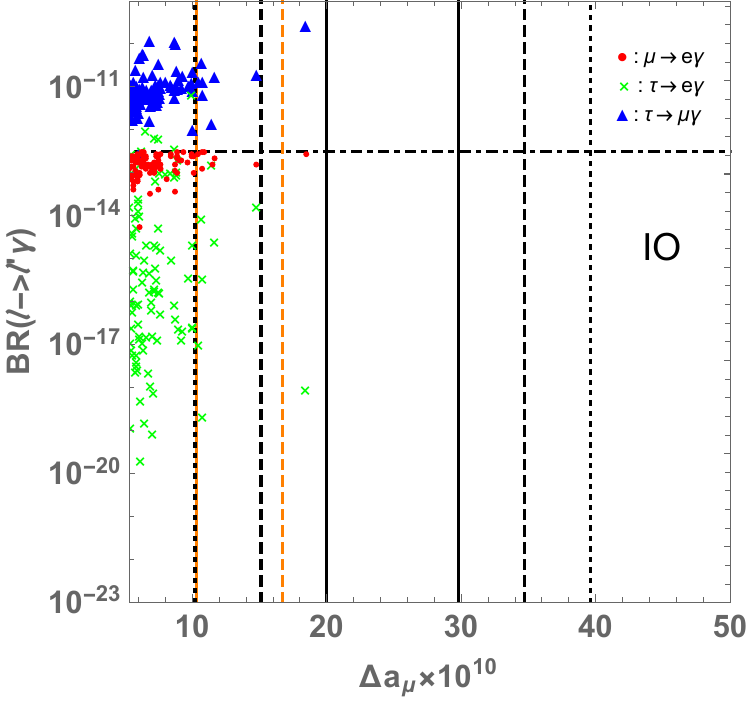}
\caption{$\Delta a_\mu$ and $BR(\ell \to \ell' \gamma)$ for allowed parameter sets where horizontal dot-dashed line shows current constraint from $BR(\mu \to e \gamma)$, and region between vertical solid, dashed and dotted lines indicate $1\sigma$, $2\sigma$ and $3\sigma$ range of  $\Delta a_\mu^{\rm data-driven}$. In addition, vertical orange-solid and orange-dashed lines indicate $1\sigma$ and $2\sigma$ upper bound of $\Delta a_\mu^{\rm WP25}$. The left(right) plot corresponds to NO(IO) case.}
\label{fig:LFV}
\end{figure}

 \begin{figure}[tb]
\includegraphics[width=7cm]{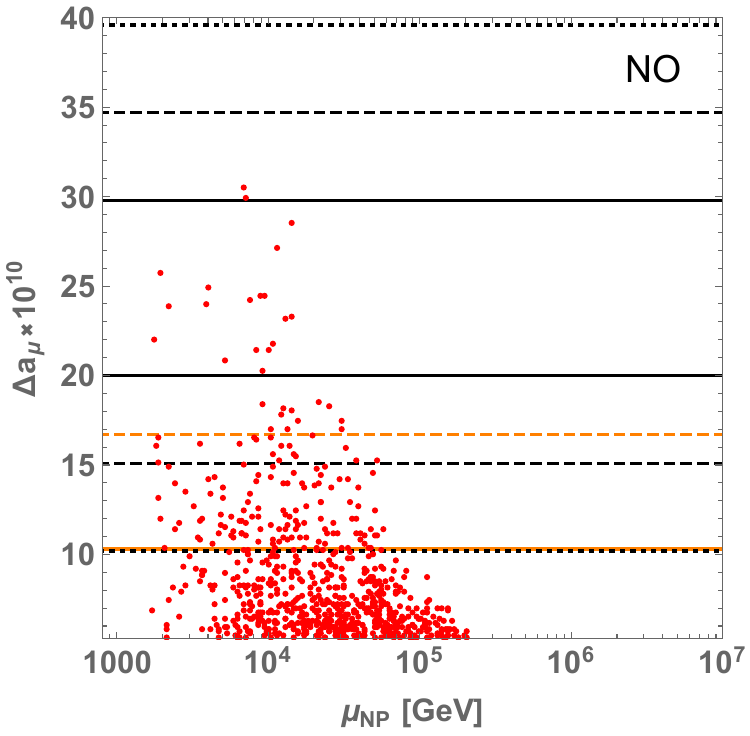}
\includegraphics[width=7cm]{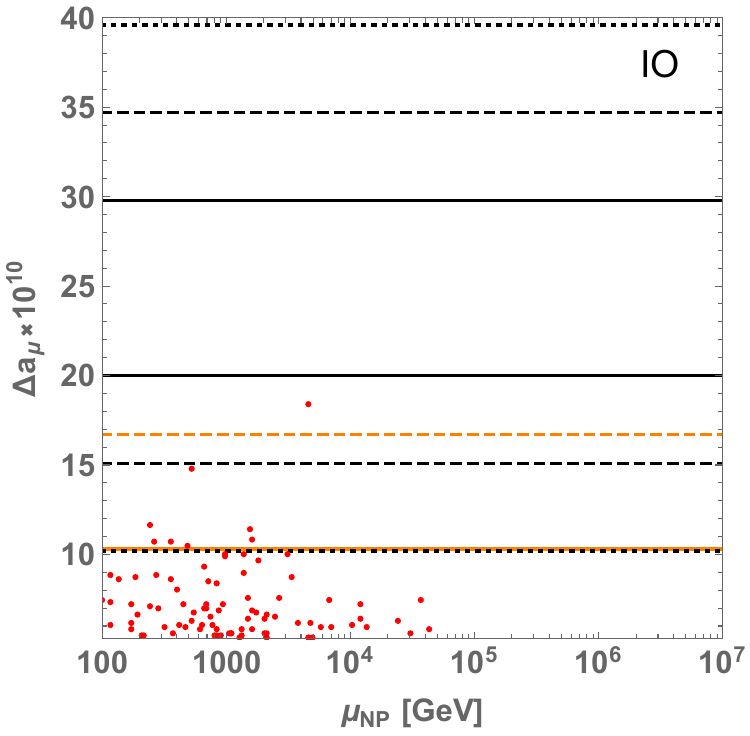}
\caption{$\mu_{NP}$ and $\Delta a_\mu$ for allowed parameter sets where region between horizontal solid, dashed and dotted lines indicate $1\sigma$, $2\sigma$ and $3\sigma$ range of $\Delta a_\mu^{\rm data-driven}$. In addition, horizontal orange-solid and orange-dashed lines indicate $1\sigma$ and $2\sigma$ upper bound of $\Delta a_\mu^{\rm WP25}$. The left(right) plot corresponds to NO(IO) case.}
\label{fig:scale}
\end{figure}

 \begin{figure}[tb]
\includegraphics[width=7cm]{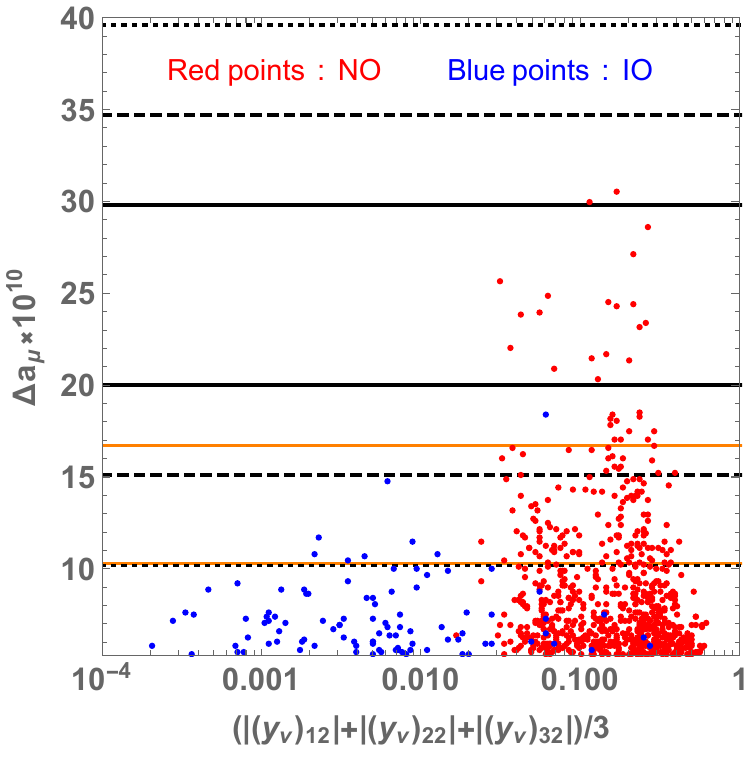}
\includegraphics[width=7cm]{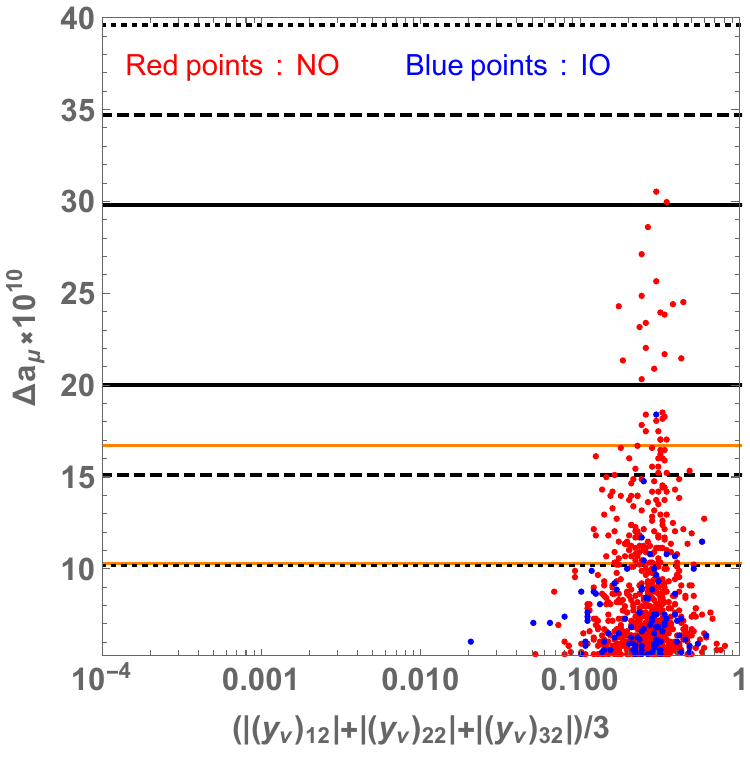}
\caption{The left(right) panel shows scatter plot for the average values of $(y_\nu)_{2k}((y_5)_{k2})$ and $\Delta a_\mu$ where the red(blue) color corresponds to NO(IO) case. The horizontal lines are the same as the previous plot.}
\label{fig:coupling}
\end{figure}

In left(right) panel of Fig.~\ref{fig:LFV}, we show the scatter plots on $\{ \Delta a_\mu, {\rm BR(\ell_i \to \ell_j \gamma)} \}$ in NO(IO) case obtained from allowed parameter sets requiring $\Delta a_\mu > 5.3 \times 10^{-10}$ as sizable value which is $4\sigma$ below the central value of  $\Delta a_\mu^{\rm data-driven}$ and allowed by $\Delta a_\mu^{\rm WP25}$ within 1 $\sigma$~\footnote{Theoretically both postive and negative $\Delta a_\mu$ is possible since the sign is determined by that of the products of Yukawa coupling, $y_\nu y_5$.}. The horizontal dot-dashed line shows current constraint from $BR(\mu \to e \gamma)$, and region between vertical solid, dashed and dotted lines indicate $1\sigma$, $2\sigma$ and $3\sigma$ range of  $\Delta a_\mu^{\rm data-driven}$. In addition, vertical orange-solid and orange-dashed lines indicate $1\sigma$ and $2\sigma$ upper bound of $\Delta a_\mu^{\rm WP25}$. The obtained values of $BR(\tau \to e(\mu) \gamma )$ are sufficiently smaller than current limits and we do not show lines corresponding to the limits. We find that IO case tends to give larger $BR(\mu \to e \gamma)$ and $BR(\tau \to e \gamma)$ values compared to NO case. Also we have less points to provide sizable muon $g-2$ in IO case since constraints from $BR(\mu \to e \gamma)$ exclude many points in this case.

In left(right) panel of Fig.~\ref{fig:scale}, we show the scatter plot on $\{\mu_{\rm NP}, \Delta a_\mu\}$ in NO(IP) case for our allowed parameter sets where region between horizontal solid, dashed and dotted lines indicate $1\sigma$, $2\sigma$ and $3\sigma$ range of muon $g-2$. We find that preferred new physics scale $\mu_{\rm NP}$ is less than $\mathcal{O}(100) (\mathcal{O}(1))$ TeV to realize sizable muon $g-2$ within $3\sigma$ of Eq.~\eqref{exp_dmu} in NO(IO) case.  In particular smaller scale is preferred to get sizable muon $g-2$ in IO case. 

 In the left(right) panel of Fig.~\ref{fig:coupling}, we show the scatter plot on the average values of Yukawa couplings $(y_\nu)_{2k}((y_5)_{k2})$ and $\Delta a_\mu$ where the red(blue) color corresponds to NO(IO) case.
We find that the values of Yukawa couplings are restricted to provide sizable muon $g-2$ while explaining neutrino mass. 
In particular the average of $(y_5)_{k2}$ should be within around $0.1$ to $1$ in both NO and IO case.

 \begin{figure}[tb]
\includegraphics[width=7cm]{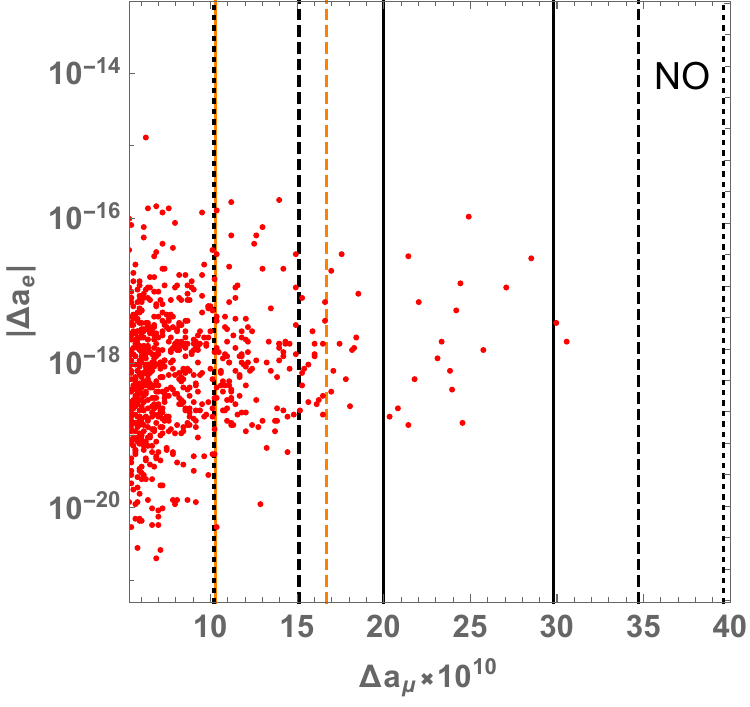}
\includegraphics[width=7cm]{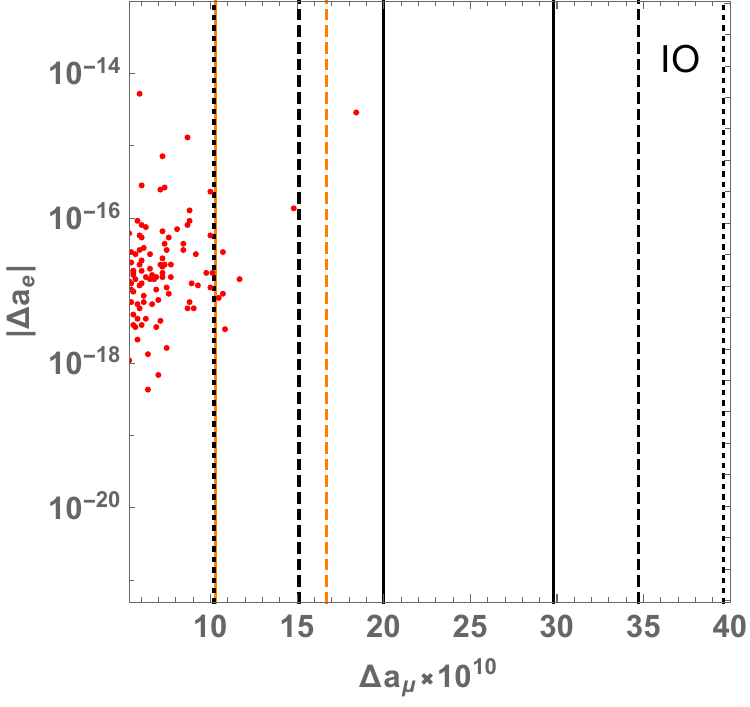}
\caption{$\Delta a_{\mu}$ and $|\Delta a_e|$ for allowed parameter sets where region between vertical solid, dashed and dotted lines indicate $1\sigma$, $2\sigma$ and $3\sigma$ range of  $\Delta a_\mu^{\rm data-driven}$. In addition, vertical orange-solid and orange-dashed lines indicate $1\sigma$ and $2\sigma$ upper bound of $\Delta a_\mu^{\rm WP25}$. The left(right) plot corresponds to NO(IO) case.}
\label{fig:g-2}
\end{figure}

 \begin{figure}[tb]
\includegraphics[width=7cm]{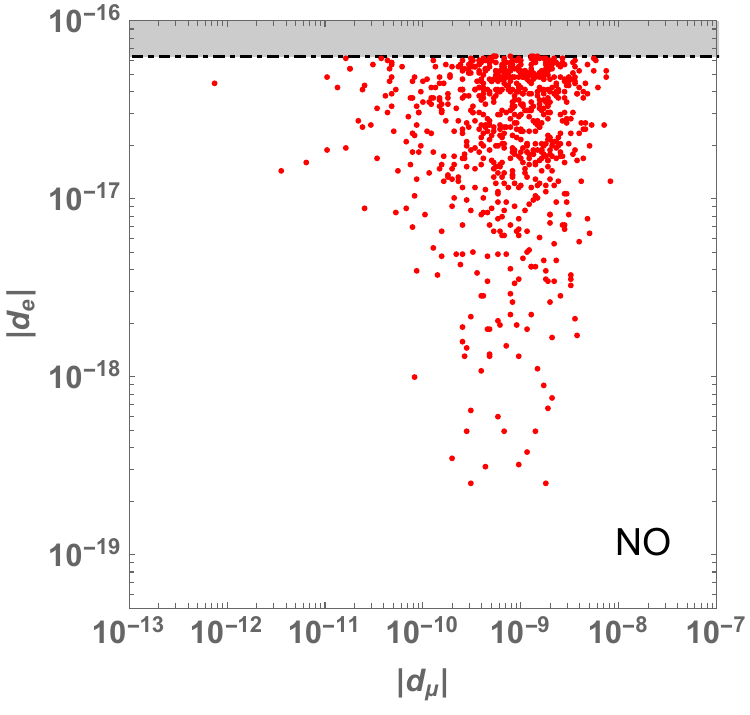}
\includegraphics[width=7cm]{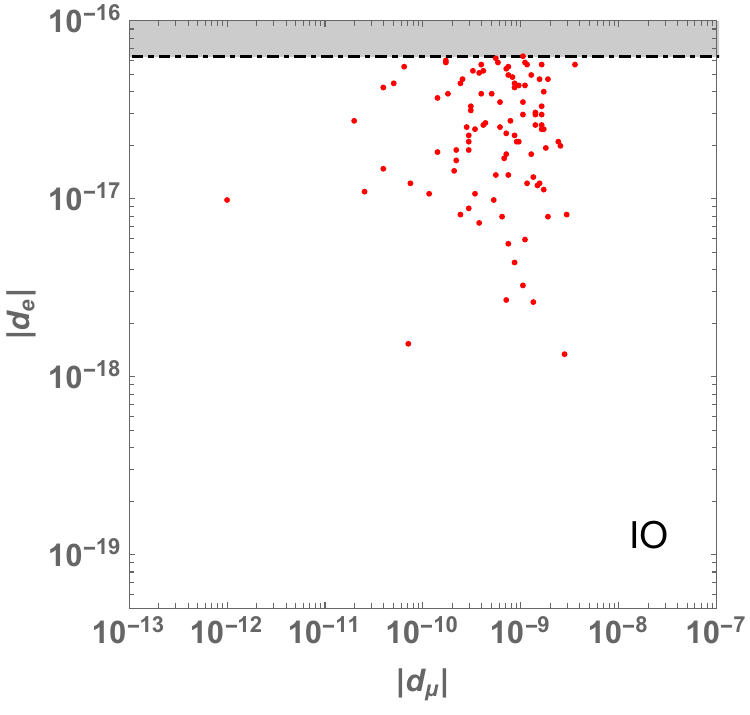}
\caption{$|d_\mu|$ and $|d_e|$ for allowed parameter sets where the gray region is excluded by experimental constraint on $|d_e|$. The left(right) plot corresponds to NO(IO) case.}
\label{fig:EDM}
\end{figure}

In left(right) panel of Fig.~\ref{fig:g-2}, we show the scatter plot on $\{\Delta a_\mu, |\Delta a_e|\}$ in NO(IO) case for our allowed parameter sets where the vertical lines are the same as Fig.~\ref{fig:LFV}. We find that the magnitude of $\Delta a_e$ is smaller than current sensitivity of it shown in Eqs.~\eqref{eq:ae-1} and \eqref{eq:ae-2} where we have same amount of points for both positive and negative sign of $\Delta a_e$. The magnitude of electron $g-2$ tends to be small since Yukawa couplings associated with electron should be small to satisfy constraint from $BR(\mu \to e \gamma)$ when we require muon $g-2$ to be $\mathcal{O}(10^{-9})$. Also we find magnitude of electron $g-2$ tends to be larger in IO case.

In left(right) panel of Fig.~\ref{fig:EDM}, we show the scatter plot on $\{|d_\mu|, |d_e|\}$ in NO(IO) case for our allowed parameter sets where the gray region is excluded by experimental constraint on $|d_e|$.  We can not find a clear correlation of them. The value of $|d_e|$ is concentrated around current limit and we expect more parameter region can be tested in future experiments.
 We also find that obtained $|d_\mu|$ is much smaller than the current upper limit. 
 
{ Finally, we discuss possible collider signatures of the model. 
In the model there are several charged particles, $\{E, H_1^\pm, H_2^\pm, H_3^\pm \}$, that can be produced via electroweak interactions at collider experiments.
For illustration let us consider mass relation $m_{E_\alpha} > m_{H_i^\pm} > m_{N_\alpha}$.
The charged scalar bosons can dominantly decay via Yukawa interaction with $y_5$ since the coupling is sizable in our scenario as 
$H^\pm \to \ell^\pm N_\alpha$ assuming $N_\alpha$ is lighter than charged scalar bosons.
The heavy neutral fermion $N$ also decays as $N \to W^\pm \ell^\mp/Z \nu$ via mixing with the SM neutrino.
Thus pair production of charged Higgs can provide signature, at $pp$-collider for example, as  
\begin{equation}
pp \to H_i^+ H_i^- \to N N \ell_1^+ \ell_2^- \to W^+ W^- \ell_1^+ \ell_2^- \ell_3^+ \ell^-_4 \ {\rm or}
 \ Z Z\ell_1^+ \ell_2^- \nu_1 \bar{\nu}_2,
\end{equation}
where the subscript of leptons just distinguish the leptons from different decay chain.
We thus expect "multi-lepton" and "multi-lepton + multi-jets" signals considering decays of the SM gauge bosons $Z$ and $W^\pm$. 
It is possible to have detectable number of events if masses of charged Higgs bosons are TeV scale at the LHC experiments.
We can also produce heavy charged lepton pairs $E^+ E^-$ via electroweak interactions.
They can decay via Yukawa interactions with coupling $y_{1,2,3,4}$ as $E^\pm \to H^\pm_i N_\alpha$ assuming it is kinematically allowed. 
Then we have the process, at $pp$ collider, such that
\begin{equation}
pp \to E^+ E^- \to N_\alpha N_\beta H_i^+ H_j^{\color{red}-},
\end{equation}
In that case we have further long decay chain processes through decays of $H^\pm_i$ and $N_\alpha$ leading many leptons and/or jets in final states.
Thus, we expect multi-leptons+jets signatures in the model along with sizable lepton $g-2$. 
To reconstruct heavy scalar/fermion masses from these signals, we need detailed simulation study and it is beyond the scope of this paper.
 }

\section{Summary and discussion}

In this work we have proposed a new inverse seesaw model based on hidden local $U(1)$ symmetry framework where 
inverse seesaw mechanism is induced at one loop level.
Two types of SM singlet chiral fermions for inverse seesaw mechanism are naturally obtained from one vector-like fermion 
where vector-like nature requires anomaly cancellation associated with hidden $U(1)$ and Majorana mass terms of $N_{L(R)}$ are forbidden at tree level.
But the Majorana mass terms
are generated at one-loop level by introducing relevant particle contents to get loop diagram.
Thus,  inverse seesaw mechanism still works well.

The same particle contents also induce contributions to lepton $g-2$/EDM and LFV decays at one loop level.
Interestingly we have obtained enhancement of them since chirality flip appears inside loop diagram picking up new heavy fermion masses.
As a result, we have obtained sizable muon $g-2$ that accommodates with deviation from the SM prediction.
We have shown LFV decay ratios, electron $g-2$ and electron/muon EDM for parameters that realizes neutrino data and sizable muon $g-2$
by performing numerical analysis.
It has been found that new physics scale is preferred to be less than $\mathcal{O}(100) (\mathcal{O}(1))$ TeV in NO(IO) case to realize muon $g-2$ with $\mathcal{O}(10^{-9})$.
The magnitude of electron $g-2$ is smaller than current central value since Yukawa coupling associated with electron should be small to avoid $BR(\mu \to e \gamma)$ when 
we require muon $g-2$ is more than $\mathcal{O}(10^{-10})$. 
Also LFV decays including electrons in final state tend to be large in IO case compared to NO case.

Notice that we obtained some viable predictions although there are many free parameters, considering some conditions.  
The mass parameters do not affect phenomenology so much such as neutrino masses and flavor physics if they are not very hierarchical, and we chose them  to be the same order as Eq.~\eqref{eq:masses} picking up new physics scale $\mu_{NP}$. Under this condition, we made global scan of free parameters $y_\nu$ and $y_5$.  
Then we obtained the results summarized in Figs.~\ref{fig:LFV}-\ref{fig:EDM} and shown some observables under requirement of $\Delta a_\mu > 5.3 \times 10^{-10}$ and neutrino mass. 
Actually we found new physics scale $\mu_{NP}$, LFVs and magnetic(electric) dipole moments which are restricted by neutrino mass and $\Delta a_\mu$ .
We expect that future improvement of precision regarding LFV decays and electron(muon) $g-2$/EDM measurements can test the model.

 In our model we have new particles and the mass scale is predicted to be $\mu_{NP} \lesssim 100$ TeV if muon $g-2$ is sizable.
Thus observation of some charged particles such as $E$ and $H^+_{1,2}$ is good test of the model that would be realized in future collider experiments.  
In fact the new charged particles can be produced via electroweak interactions and it would provide us multi-lepton + jets signals at collider experiments
where further detailed analysis is left for future works. 

\section*{Acknowledgments}
T.~N. was supported by the Fundamental Research Funds for the Central Universities.
H.~O. is supported by Zhongyuan Talent (Talent Recruitment Series) Foreign Experts Project. 

\begin{appendix}

\section{Expressions of subdominant one-loop contribution for $Z \to \ell_i \bar{\ell}_j $}

Here we summarize expressions for $\delta A_{R(L)}$, $\delta B_{R(L)}$ and $\delta C_{R(L)}$ as follows:
\begin{align}
(\delta A_{R})_{ij} \simeq & -\frac{(y_5^\dagger)_{i \alpha} (y^\dagger_\nu)_{\alpha j}}{(4 \pi)^2 s_W^2}   M_{E_\alpha} m_i s_\beta \int [dX] (1+2x-2z)  \nonumber \\
& \times \left[ \frac{c_A s_B c_B }{\Delta_{H_{11}}} \left( \left( \frac12 - s_W^2  \right) c_B^2 - s_W^2 c_A^2 s_B^2 \right)  
+ \frac{c_A s_B^2  }{\Delta_{H_{12}}} \left( \left( \frac12 - s_W^2  \right) c_B s_B + s_W^2 c_A^2 s_B c_B \right) \right.  \nonumber \\ 
& \left. \quad  - \frac{c_A c_B^2  }{\Delta_{H_{21}}} \left( \left( \frac12 - s_W^2  \right) c_B s_B + s_W^2 c_A^2 s_B c_B \right) 
- \frac{c_A s_B c_B  }{\Delta_{H_{22}}} \left( \left( \frac12 - s_W^2  \right) s_B^2 - s_W^2 c_A^2 c_B^2 \right)  \right], \\  
(\delta A_{L})_{ij} \simeq & \frac{(y_5^\dagger)_{i \alpha} (y^\dagger_\nu)_{\alpha j}}{(4 \pi)^2 } \left( -\frac12 + s^2_W \right)^{-1}   M_{E_\alpha} m_j s_\beta \int [dX] (1-2z)  \nonumber \\
& \times \left[ \frac{c_A s_B c_B }{\Delta_{H_{11}}} \left( \left( \frac12 - s_W^2  \right) c_B^2 - s_W^2 c_A^2 s_B^2 \right)  
+ \frac{c_A s_B^2  }{\Delta_{H_{12}}} \left( \left( \frac12 - s_W^2  \right) c_B s_B + s_W^2 c_A^2 s_B c_B \right) \right.  \nonumber \\ 
& \left. \quad  - \frac{c_A c_B^2  }{\Delta_{H_{21}}} \left( \left( \frac12 - s_W^2  \right) c_B s_B + s_W^2 c_A^2 s_B c_B \right) 
- \frac{c_A s_B c_B  }{\Delta_{H_{22}}} \left( \left( \frac12 - s_W^2  \right) s_B^2 - s_W^2 c_A^2 c_B^2 \right)  \right],
\end{align}
where $\Delta_{H_{ab}} \simeq -z(1-z+x) m_Z^2 + x M^2_{E_\alpha} + y m^2_{H_a^+} + z m^2_{H_b^+}$ ignoring charged lepton mass.
\begin{align}
(\delta B_{R})_{ij} \simeq & \frac{(y_\nu)_{i \alpha} (y_5)_{\alpha j}}{(4 \pi)^2 s_W^2}   M_{E_\alpha} m_j s_\beta \int [dX] (1-2z)  \nonumber \\
& \times \left[ \frac{c_A s_B c_B }{\Delta_{H_{11}}} \left( \left( \frac12 - s_W^2  \right) c_B^2 - s_W^2 c_A^2 s_B^2 \right)  
+ \frac{c_A s_B^2  }{\Delta_{H_{12}}} \left( \left( \frac12 - s_W^2  \right) c_B s_B + s_W^2 c_A^2 s_B c_B \right) \right.  \nonumber \\ 
& \left. \quad  - \frac{c_A c_B^2  }{\Delta_{H_{21}}} \left( \left( \frac12 - s_W^2  \right) c_B s_B + s_W^2 c_A^2 s_B c_B \right) 
- \frac{c_A s_B c_B  }{\Delta_{H_{22}}} \left( \left( \frac12 - s_W^2  \right) s_B^2 - s_W^2 c_A^2 c_B^2 \right)  \right], \\  
(\delta B_{L})_{ij} \simeq & -\frac{(y_\nu)_{i \alpha} (y_5)_{\alpha j}}{(4 \pi)^2 } \left( -\frac12 + s^2_W \right)^{-1}   M_{E_\alpha} m_i s_\beta \int [dX] (1+2x-2z)  \nonumber \\
& \times \left[ \frac{c_A s_B c_B }{\Delta_{H_{11}}} \left( \left( \frac12 - s_W^2  \right) c_B^2 - s_W^2 c_A^2 s_B^2 \right)  
+ \frac{c_A s_B^2  }{\Delta_{H_{12}}} \left( \left( \frac12 - s_W^2  \right) c_B s_B + s_W^2 c_A^2 s_B c_B \right) \right.  \nonumber \\ 
& \left. \quad  - \frac{c_A c_B^2  }{\Delta_{H_{21}}} \left( \left( \frac12 - s_W^2  \right) c_B s_B + s_W^2 c_A^2 s_B c_B \right) 
- \frac{c_A s_B c_B  }{\Delta_{H_{22}}} \left( \left( \frac12 - s_W^2  \right) s_B^2 - s_W^2 c_A^2 c_B^2 \right)  \right],
\end{align}
\begin{align}
(\delta C_{R})_{ij} \simeq & \ - \frac{c_A c_B s_B s_\beta}{(4\pi)^2} \left[ m_i (y_\nu)_{i \alpha} (y_5)_{\alpha j} + m_j y_5^\dagger)_{i \alpha} (y^\dagger_\nu)_{\alpha j} \right] \nonumber \\
& \times M_{E_\alpha} \int [dX]_2 y(1-y) \left[ \frac{1}{M^2_{E_\alpha} x + m^2_{H^+_1} y} - \frac{1}{M^2_{E_\alpha} x + m^2_{H^+_2} y} \right], \nonumber \\
(\delta C_{L})_{ij} \simeq & \ - \frac{c_A c_B s_B s_\beta}{(4\pi)^2} \left[ m_j (y_\nu)_{i \alpha} (y_5)_{\alpha j} + m_i y_5^\dagger)_{i \alpha} (y^\dagger_\nu)_{\alpha j} \right] \nonumber \\
& \times M_{E_\alpha} \int [dX]_2 y(1-y) \left[ \frac{1}{M^2_{E_\alpha} x + m^2_{H^+_1} y} - \frac{1}{M^2_{E_\alpha} x + m^2_{H^+_2} y} \right],
\end{align}
where $\int [dX]_2 = \int_0^1 dx dy \delta(1-x-y)$.

\end{appendix}

\bibliography{U1Inverse.bib}
\end{document}